\documentclass[twocolumn,aps]{revtex4}
\usepackage{amsmath}
\usepackage{amssymb}
\usepackage{graphicx}
\usepackage{dcolumn}
\usepackage{bm}
\usepackage{xcolor}
\usepackage{epstopdf}
\usepackage{longtable}

\begin{document}

\preprint{Phys.Rev.B }

\title{Hydrodynamic magnetotransport in a GaAs Corbino geometry}
\author{A. D. Levin,$^1$   A. S. Jaroshevich,$^{2}$ Z. D. Kvon,$^{2,3}$  V. A. Chitta,$^1$ M. S. Aksenov, $^{2}$  D. V. Dmitriev,$^{2}$ A. K. Bakarov$^{2}$ and G. M. Gusev$^1$}

\affiliation{$^1$Instituto de F\'{\i}sica da Universidade de S\~ao
Paulo, 135960-170, S\~ao Paulo, SP, Brazil}
\affiliation{$^2$Institute of Semiconductor Physics, Novosibirsk
630090, Russia}
\affiliation{$^3$Novosibirsk State University, Novosibirsk 630090,
Russia}

\date{\today}
\begin{abstract}
We report the observation of positive magnetoresistance in high-mobility GaAs Corbino devices. Over a broad intermediate-temperature range, the resistance exhibits a quadratic dependence on perpendicular magnetic field. We analyze the data within hydrodynamic theories of magnetotransport in the Corbino geometry, which describe the crossover between the diffusive and viscous regimes,
including finite-slip boundary conditions appropriate for current-penetrable contacts.
The extracted relaxation rates are consistent with an approximately $T^2$ temperature dependence of the electron-electron scattering contribution.
The extracted viscous relaxation parameters are consistent with those obtained  from Hall-bar measurements.
Comparison with theory indicates that the observed magnetoresistance is predominantly governed by the bulk hydrodynamic response,
while finite slip modifies the Stokes--Ohm crossover quantitatively
and the field-dependent boundary voltage provides a separate correction.
These results show that Corbino magnetotransport can serve as a complementary bulk-sensitive probe of viscous electron flow.
\end{abstract}
\maketitle
\section{Introduction}

Charge transport in ordinary conductors is usually described within the Drude picture, where momentum-relaxing scattering by disorder and phonons determines the electrical response. A qualitatively different regime emerges when momentum-conserving electron-electron collisions occur more frequently than momentum-relaxing events. In that case, momentum is efficiently redistributed within the electron system before being lost to the lattice, and the carrier dynamics acquires a collective character that is more naturally described by hydrodynamics than by single-particle kinetics \cite{narozhny,hui,gurzhi,andreev}. For many years this regime remained difficult to access experimentally, since it requires exceptionally clean materials and an appropriate temperature window. The development of ultrahigh-mobility two-dimensional electron systems has made such conditions experimentally achievable.

Once the hydrodynamic regime is established, transport displays a variety of phenomena that have no simple counterpart in conventional diffusive conduction. These include Gurzhi effect \cite{gurzhi, dejong, gusev1,gusev2}, negative nonlocal resistance \cite{bandurin, levin}, electron whirlpools \cite{palm} , Poiseuille-like flow \cite{sulpizio, patricio}, superballistic transport through point contacts \cite{kumar, ginzburg, pogosov}, Stokes-like flow around obstacles \cite{lucas, gusev3, levin2, krebs}, the electronic Venturi effect, in which the interaction between non-equilibrium and degenerate carriers drives a directed electron jet \cite{taubert} and departures from the Wiedemann-Franz law \cite{principi2, sarma, pusep2}. In a perpendicular magnetic field, the phenomenology becomes even richer and includes negative magnetoresistance \cite{alekseev, raichev2, gusev1}, Hall-viscous effects \cite{scaffidi, gromov, berdyugin, gusev4}, and geometry-dependent viscous magnetotransport \cite{moessner, keser, levin3}. Together, these observations establish electron hydrodynamics as a distinct transport regime of interacting mesoscopic systems.

The device geometry plays a central role in identifying hydrodynamic effects. Most experiments have been performed in Hall-bar structures, where the interpretation is often complicated by the presence of side boundaries. In magnetic field, the measured resistance may depend not only on the bulk viscosity, but also on the magnetic-field dependence of viscous coefficients, the relation between the cyclotron radius and the channel width, edge-current contributions, and the details of the boundary condition at the sample edges \cite{keser, moessner, gusev4,levin3}. Since these boundary conditions are rarely known microscopically, they introduce a substantial uncertainty into the analysis. Considerable effort has therefore been devoted to engineering channels with more controllable boundary properties and to designing geometries in which viscous signatures are enhanced in a predictable manner \cite{keser, levin3}.

The Corbino geometry provides an attractive alternative. Because a Corbino disk has no extended side edges, it is naturally suited for probing bulk transport \cite{falkovich, kumar2, levchenko, tomadin, afrose, gervais, gervais2, levchenko2, raichev}. In the hydrodynamic context, the Corbino geometry is especially appealing because a perpendicular magnetic field converts viscous flow into a measurable magnetoresistance signal. In a doped two-dimensional electron gas, theory predicts a positive magnetoresistance quadratic in magnetic field and directly linked to viscous transport in the Corbino geometry \cite{levchenko,levchenko2}. At lower carrier density or in the presence of substantial thermal gradients, additional contributions may arise, but in the high-density Fermi-liquid regime the Corbino disk remains one of the cleanest platforms for studying bulk hydrodynamic magnetotransport. 
At the same time, the absence of extended side edges does not eliminate the role of the source and drain interfaces. For current-penetrable contacts, the tangential flow is generally described by a finite-slip boundary condition, while the electrochemical potential may exhibit a discontinuity across the contact interface \cite{raichev}. In a perpendicular magnetic field, this potential jump acquires an additional field-dependent contribution, providing a boundary correction to the bulk Corbino magnetoresistance.

From the hydrodynamic point of view, the Corbino geometry is also unusual. For the essentially potential radial flow in a Corbino disk, the divergence of the viscous stress tensor vanishes in the bulk. This can be understood intuitively: the viscous force density is governed by spatial variations of the vorticity, and the radial Corbino flow is irrotational (potential), so its vorticity---and hence the viscous force---vanishes in the bulk, even though the velocity-gradient (shear) stress itself remains finite. As a result, the viscous force density is zero even though the flow is not dissipationless. In linear response, where the total force density must vanish, the external driving force is therefore effectively expelled from the interior of the sample \cite{falkovich, kumar2}. In electron fluids, this force expulsion appears as voltage and temperature drops localized near the boundaries, with amplitudes governed by the viscosity. At the same time, energy dissipation still occurs in the bulk because the viscous stress tensor itself remains finite even when its divergence vanishes. Thus, in Corbino flow, viscosity manifests itself through the stress distribution and the associated energy loss rather than through a finite bulk viscous force.

Recent experiments on Corbino devices have demonstrated that the contact regions can support viscous boundary layers \cite{kumar2}. In high-mobility two-dimensional systems, the conductivity mismatch between metallic leads and the electron fluid, together with microscopic roughness at the interfaces, tends to make the injected current nearly normal to the contacts. The source and drain terminals therefore become special regions where the hydrodynamic flow adjusts to the boundary conditions and where additional resistance contributions may arise \cite{falkovich}. 
A microscopic treatment of such current-penetrable
interfaces shows that their influence can be characterized by a slip
length together with a boundary condition for the electrochemical-potential
jump \cite{raichev}. The slip length $l_s$ characterizes how freely the
tangential electron flow moves along the interface: $l_s=0$ corresponds
to the no-slip limit, whereas a finite $l_s$ allows a nonzero tangential
current at the boundary. Consequently, finite slip can quantitatively
modify the Stokes--Ohm crossover. In a perpendicular magnetic field, the
electrochemical-potential jump acquires an additional field-dependent
contribution, providing a separate boundary correction to the bulk
Corbino magnetoresistance.

These considerations motivate the present work. We investigate magnetotransport in high-mobility GaAs Corbino devices and analyze the results within hydrodynamic theories appropriate to the Corbino geometry. Our main goal is to identify the origin of the observed positive magnetoresistance and to separate the dominant bulk contribution from possible boundary-related corrections. In the interior of the disk, away from the contacts, the current approaches a Drude-like profile governed by carrier density and momentum relaxation. Near the contacts, however, viscous boundary layers may form and modify the current distribution. 
We therefore extend the Stokes--Ohm crossover analysis to finite-slip boundary conditions and use the resulting crossover function to extract the viscous transport parameters.
The Corbino geometry therefore provides a natural setting for testing hydrodynamic theories of magnetoresistance while minimizing the ambiguities associated with extended sample edges.

\begin{figure} 
 \includegraphics[width=8cm]{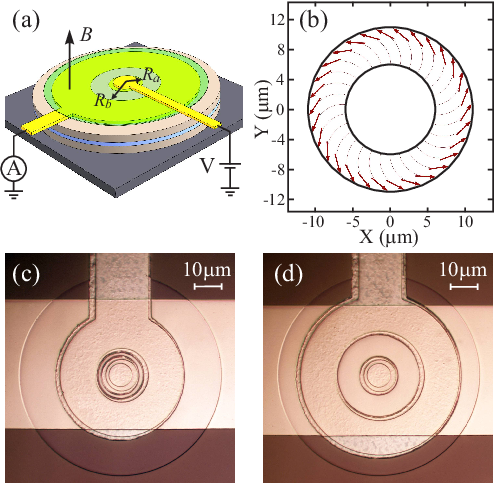}
\caption{(Color online) (a) Schematic of the Corbino disk geometry. (b) Hydrodynamic velocity-flow profile in a Corbino device at low magnetic field. (c) Optical image of the small-ring sample. (d) Optical image of the large-ring sample.}
\end{figure}

In this paper we show that GaAs Corbino devices exhibit a pronounced positive magnetoresistance with an approximately quadratic dependence on perpendicular magnetic field over a broad intermediate-temperature range. We compare the data with hydrodynamic theories describing the crossover between Ohmic and viscous regimes in the Corbino geometry 
including finite-slip boundary conditions at the current-penetrable contacts.
Our analysis indicates that the dominant contribution to the magnetoresistance originates from bulk hydrodynamic transport, while 
finite slip modifies the crossover quantitatively and the
field-dependent contact contribution provides a separate correction.

\section{EXPERIMENTAL RESULTS}
\begin{figure*} 
 \includegraphics[width=18cm]{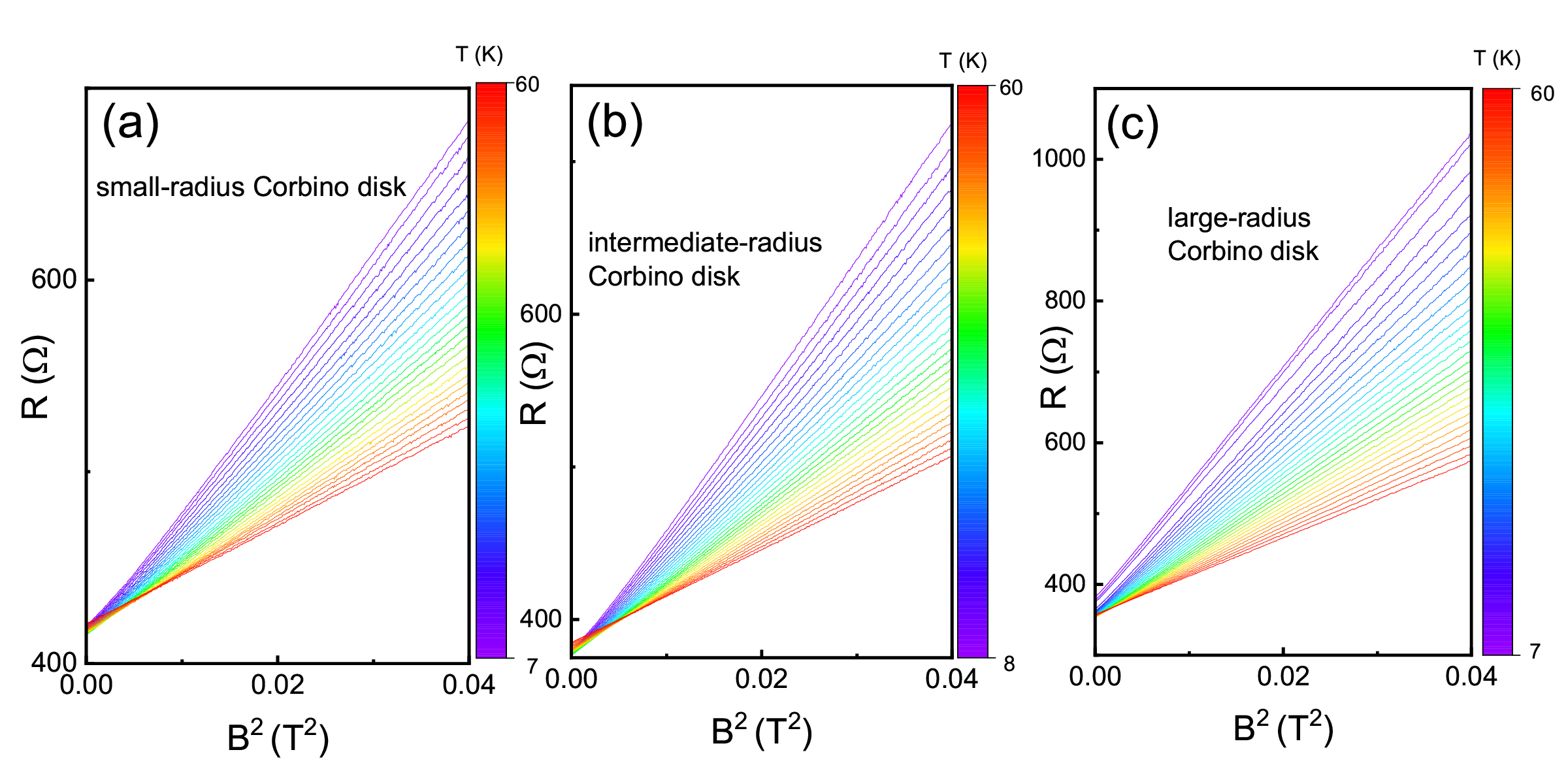}
\caption{(Color online) (a) Magnetoresistance of the small radius ($R_b=11 \mu m$) Corbino disk as a function of the temperature. (b) Magnetoresistance of the intermediate radius ($R_b=12 \mu m$) Corbino disk as a function of the temperature. (c) Magnetoresistance of the large radius ($R_b=18 \mu m$) Corbino disk as a function of the temperature }
\end{figure*}
We used high-quality GaAs quantum wells for the fabrication of our devices. The quantum wells had a width of \(14~\mathrm{nm}\) and an electron density  of approximately $n_s=7.1\times10^{11}~\mathrm{cm^{-2}}$ at \(T=4.2~\mathrm{K}\). The low-temperature mobility was \(\mu=2\times10^{6}~\mathrm{cm^2/Vs}\).

For the present electron density, the Shubnikov–de Haas oscillations measured in the macroscopic Hall-bar samples fabricated from the same wafer show a single oscillation frequency, with no signature of second-subband occupation.

The measurements were performed using Corbino devices. Each Corbino device consists of concentric inner and outer ohmic contacts, which define an annular current path in the two-dimensional electron gas. The inner radius was fixed at \(R_a=6~\mu\mathrm{m}\) for all devices, while the outer radius \(R_b\) was varied to realize three different geometries: small-radius, intermediate-radius, and large-radius Corbino disks with \(R_b=11~\mu\mathrm{m}\), \(12~\mu\mathrm{m}\), and \(18~\mu\mathrm{m}\), respectively.
Optical images and schematics of the Corbino disks are shown in Fig.~1.
The Corbino structures were covered with a \(140~\mathrm{nm}\)-thick \(\mathrm{SiO_2}\) dielectric layer. The metal lead to the central contact was evaporated on top of this dielectric layer in order to electrically isolate it from the outer contact and prevent a short circuit between the inner and outer electrodes. Ohmic contacts to the two-dimensional electron gas were fabricated by annealing Ti/Ni/Au metallization deposited on the GaAs surface.

The contact resistance of the Corbino devices was relatively large,
typically \(300\)--\(400~\Omega\), considerably exceeding the intrinsic
resistance of the two-dimensional electron gas in the annulus. The
corresponding bulk sheet resistivity is
\(\rho_0=1/(n_s e\mu_0)\sim4~\Omega\), while the resistance of the
annular 2DES itself is substantially smaller. This is an inherent
disadvantage of the two-terminal Corbino geometry compared with
multiterminal Hall-bar measurements, where the series contact
contribution is excluded from the measured longitudinal resistance.

We note that multiterminal Corbino geometries, such as
those employed in Refs.~\cite{gervais,gervais2}, provide an alternative
way to reduce the direct contribution of the current contacts to the
measured voltage. However, these structures contain additional
concentric metallic contacts and therefore introduce additional
current-penetrable boundaries for the electron fluid. In the
hydrodynamic regime, each such interface may be associated with finite
slip and an electrochemical-potential discontinuity, so that the
multiterminal geometry corresponds to a different, multiboundary
hydrodynamic problem rather than simply to a four-terminal version of
the single-annulus geometry considered here.

In the present two-terminal geometry, a
field-independent series contact resistance cancels from
\(R(B)-R(0)\) and therefore does not enter the bulk magnetoresistance
prefactor. Contact effects can nevertheless contribute through the
finite-slip boundary condition and through the magnetic-field-dependent
electrochemical-potential jump at current-penetrable interfaces, as
discussed below.
Nevertheless, the current-voltage characteristics remained linear both
at zero and finite magnetic field, confirming ohmic operation of the
contacts over the measurement range.

Transport measurements were performed in a VTI cryostat using direct voltage measurements. To avoid electron overheating, a dc current of \(1~\mu\mathrm{A}\) was applied through the sample, which was sufficiently low to remain in the linear-response regime. In parallel, we measured Hall-bar devices fabricated from the same wafer, including devices used in our previous work~\cite{gusev1, levin2, levin3}. These measurements were used to independently determine the parameters of the two-dimensional electron gas in zero and finite magnetic fields.

Although the contact resistance in the Corbino devices was almost two orders of magnitude larger than the bulk resistance of the two-dimensional electron gas, the magnetoresistance signal was large and could be extracted unambiguously from the measured field dependence.

Figure~2 shows the magnetoresistance of Corbino disks with three different outer radii, measured at different temperatures and plotted as a function of \(B^2\). These data demonstrate that, over the entire low-field range used in our measurements, up to \(B=0.2~\mathrm{T}\), the magnetoresistance follows a purely quadratic magnetic-field dependence. At zero magnetic field, the resistance exhibits a nonmonotonic temperature dependence. At low temperatures, the resistance decreases slightly with increasing temperature, which is qualitatively consistent with the Gurzhi effect. At higher temperatures, the resistance shows a weak increase with temperature, which we attribute to phonon scattering~\cite{gusev1,levin2,levin3}.

Despite the relatively large contact resistance, the magnetoresistance signal is clearly resolved and has a large magnitude. Assuming that the resistance of the two-dimensional electron gas in the Corbino annulus is comparable to that of the bulk Hall-bar samples$ ~R_{0}=\frac{\rho_0}{2\pi}ln\frac{R_b}{R_a}\approx 0.5-1$ Ohm, the relative magnetoresistance can be estimated to be as large as \(\Delta R/R_{0}\sim 10^4\%\). Comparing devices with different outer radii, we find that the magnetoresistance increases approximately linearly with \(R_b\). In addition, the magnetoresistance decreases strongly with increasing temperature.

Thus, in the full magnetic-field range studied here, the magnetoresistance can be described by
\[
R(B)-R(0)=\alpha B^2,
\]
where the coefficient \(\alpha\) depends on both the device radius and temperature.
\begin{figure} 
 \includegraphics[width=9cm]{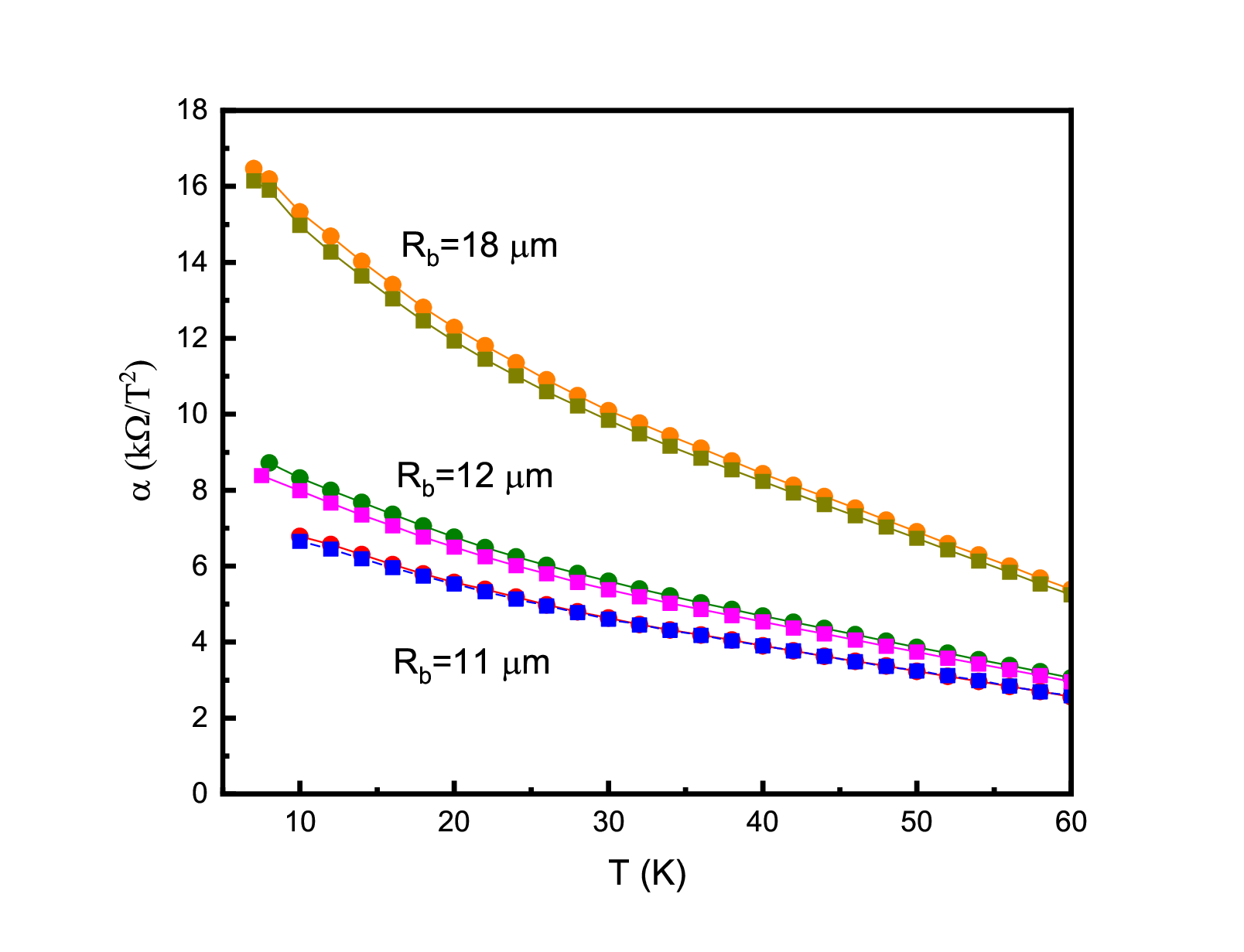}
\caption{(Color online) Temperature dependence of the quadratic magnetoresistance coefficient \(\alpha\) for Corbino devices with different outer radii. Two samples were studied for each device radius.}
\end{figure}

Figure~3 shows the temperature dependence of the quadratic magnetoresistance coefficient \(\alpha\) for Corbino devices with different outer radii. For each device geometry, measurements were performed on two nominally identical samples. The coefficient \(\alpha\) decreases strongly with increasing temperature. In addition, \(\alpha\) becomes smaller as the outer radius \(R_b\) is reduced. To clarify the origin of this behavior, we compare the experimental results with theoretical models in the following section.

Although a weak magnetic-field dependence of the contact contribution cannot be completely ruled out in a two-terminal Corbino geometry, several observations argue against a contact-dominated interpretation of the measured quadratic magnetoresistance. The measurements were performed in the weak-field regime, the coefficient \(\alpha\) is reproducible across nominally identical samples, and its temperature and radius dependences are reasonably captured by the bulk crossover model \cite{levchenko2}, considered below. In addition, \(\alpha\) was found to be independent of the applied current within the investigated range, indicating that the observed response is not caused by trivial contact-related nonlinearities. These observations demonstrate that the quadratic response is robust
and reproducible, but they do not by themselves exclude a
magnetic-field-dependent boundary contribution. We therefore treat
this contribution as a separate systematic correction, as discussed
below.
\section{Theoretical model}

The interpretation of magnetoresistance in the Corbino geometry is less straightforward than in Hall-bar devices. In a Corbino disk, the classical Drude response already produces a quadratic $B^2$ magnetoresistance, so the observation of positive magnetoresistance does not by itself provide unambiguous evidence for hydrodynamic transport. This differs from the Hall-bar geometry, where the classical Drude longitudinal magnetoresistance vanishes and the observed negative magnetoresistance can be linked much more directly to viscous flow \cite{alekseev}. We therefore analyze the present data not as a standalone proof of hydrodynamics, but within the framework of earlier work that has already established hydrodynamic corrections in high-mobility two-dimensional electron systems \cite{dejong, gusev1, gusev2, levin, levin2}.

The relevant hydrodynamic length scale is the Gurzhi length, $l_G=\sqrt{\nu\tau_{\mathrm{mr}}}$, where $\nu$ is the kinematic viscosity and $\tau_{\mathrm{mr}}$ is the momentum-relaxation time. The kinematic viscosity is related to the relaxation time of the second angular harmonic of the distribution function, $\tau_2$, by $\nu=\frac{v_F^2\tau_2}{4}$,so that $l_G=\frac{v_F}{2}\sqrt{\tau_2\tau_{\mathrm{mr}}}$.

We start from the Stokes--Ohm description of electron flow in an annular geometry \cite{levchenko2}. A perpendicular magnetic field generates an azimuthal component of the current, whose viscous relaxation modifies the radial electric field and produces an additional positive magnetoresistance. In contrast to the ideal no-slip problem, however, the metallic contacts of a Corbino disk are current-penetrable interfaces. We therefore extend the Stokes--Ohm crossover solution to finite-slip boundary conditions, following the boundary formulation introduced for current-penetrable contacts in Ref.~\cite{raichev}.

For the bulk contribution we write
\begin{equation}
R_{\mathrm{bulk}}(B,T)-R_{\mathrm{bulk}}(0,T)
=
\frac{B^2\ln a}
{2\pi\rho_0(T)(n_s e)^2}
\,
\Phi
\left(
a,b,\lambda
\right),
\label{eq:bulkMR}
\end{equation}
where
\[
a=\frac{R_b}{R_a},
\qquad
b=\frac{R_b}{l_G},
\qquad
\lambda=\frac{l_s}{l_G},
\]
and $l_s$ is the slip length. The quantity $\rho_0(T)$ is the intrinsic bulk Drude sheet resistivity,
\begin{equation}
\rho_0(T)=
\frac{m^\ast}
{n_s e^2\tau_{\mathrm{mr}}(T)}.
\label{eq:rho0}
\end{equation}
Here and below the same momentum-relaxation time $\tau_{\mathrm{mr}}(T)$ enters both $\rho_0(T)$ and the Gurzhi length $l_G(T)$.

 We define $\Phi$ directly as the physical crossover function. It satisfies $\Phi\rightarrow0$ \text{in the Stokes limit},
and $\Phi\rightarrow1$ \text{in the Ohmic limit}. The azimuthal current obtained from the Stokes--Ohm equation can be written in the form
\begin{equation}
j_\phi(r)=
\frac{IB}
{2\pi\rho_0 n_s e\,l_G}
\left[
C_I I_1(x)+C_KK_1(x)-\frac{1}{x}
\right],
\end{equation}
where $x=\frac{r}{l_G}$, and $I_n$ and $K_n$ are modified Bessel functions.

For finite slip, the tangential current satisfies Navier boundary conditions at the inner and outer contacts,
\begin{equation}
j_\phi(R_a)
=
l_s
\left.
\frac{\partial j_\phi}{\partial r}
\right|_{R_a},
\qquad
j_\phi(R_b)
=
-l_s
\left.
\frac{\partial j_\phi}{\partial r}
\right|_{R_b}.
\label{eq:slipBC}
\end{equation}

Introducing $x_a=\frac{b}{a}, x_b=b$, the coefficients $C_I$ and $C_K$ are determined from
\begin{widetext}
\begin{equation}
\begin{pmatrix}
I_1(x_a)-\lambda I_1'(x_a) &
K_1(x_a)-\lambda K_1'(x_a)
\\
I_1(x_b)+\lambda I_1'(x_b) &
K_1(x_b)+\lambda K_1'(x_b)
\end{pmatrix}
\begin{pmatrix}
C_I\\
C_K
\end{pmatrix}
=
\begin{pmatrix}
x_a^{-1}+\lambda x_a^{-2}
\\
x_b^{-1}-\lambda x_b^{-2}
\end{pmatrix}.
\label{eq:matrix}
\end{equation}
\end{widetext}
The resulting finite-slip crossover function is
\begin{equation}
\Phi(a,b,\lambda)
=
1-
\frac{
C_I\left[I_0(b)-I_0(b/a)\right]
+
C_K\left[K_0(b/a)-K_0(b)\right]
}
{\ln a}.
\label{eq:phiSlip}
\end{equation}

For $l_s=0$, Eq.~(\ref{eq:phiSlip}) reduces to the no-slip Stokes--Ohm crossover of Ref.~\cite{levchenko2}. Finite $l_s$ increases the penetration of the azimuthal flow into the contact region and therefore quantitatively shifts the crossover between the viscous and Ohmic regimes.

An important check of Eq.~(\ref{eq:phiSlip}) is provided by its two limiting cases. In the purely diffusive limit, $b=\frac{R_b}{l_G}\gg1$, the viscous boundary layers become narrow compared with the sample dimensions and $\Phi(a,b,\lambda)\rightarrow1$.

Equation~(\ref{eq:bulkMR}) then reduces to the conventional Drude Corbino magnetoresistance,
\begin{equation}
\Delta R_{\mathrm{diff}}
=
\frac{B^2\ln a}
{2\pi\rho_0(n_s e)^2}.
\label{eq:diffLimit}
\end{equation}
Equivalently,
\begin{equation}
\Delta R_{\mathrm{diff}}
=
R_0(\mu B)^2,
\end{equation}
where $R_0=\frac{\rho_0}{2\pi}\ln a$ is the zero-field bulk resistance of the annulus and
$\mu=e\tau_{\mathrm{mr}}/m^\ast$.

In the opposite, purely hydrodynamic or Stokes limit, $b\ll1$, the crossover function behaves as
\begin{equation}
\Phi(a,b,\lambda)
\simeq
\frac{b^2(1-a^{-2})}
{8\ln a}
G_{\mathrm{corb}},
\label{eq:stokesPhi}
\end{equation}
where $G_{\mathrm{corb}}$ is the finite-slip geometrical factor introduced in Ref.~\cite{raichev}. Using
$l_G^2=\nu\tau_{\mathrm{mr}}$ and
$\rho_0=m^\ast/(n_s e^2\tau_{\mathrm{mr}})$,
the momentum-relaxation time cancels explicitly, and Eq.~(\ref{eq:bulkMR}) becomes
\begin{equation}
\Delta R_{\mathrm{hyd}}
=
\frac{B^2(R_b^2-R_a^2)}
{16\pi\eta}
G_{\mathrm{corb}},
\label{eq:hydroLimit}
\end{equation}
where $\eta=m^\ast n_s\nu$ is the dynamic shear viscosity.

Equation~(\ref{eq:hydroLimit}) demonstrates an important distinction between the two limits. In the purely hydrodynamic regime the magnetoresistance is independent of the momentum-relaxing Drude resistivity and is controlled directly by the viscosity. In the purely diffusive regime, Eq.~(\ref{eq:diffLimit}), the viscous contribution disappears and the response is completely determined by ordinary momentum relaxation. The full finite-slip expression, Eq.~(\ref{eq:phiSlip}), continuously connects these two limits without introducing additional interpolation parameters.

The finite-slip crossover is also consistent with the Stokes-limit result derived independently for current-penetrable contacts in Ref.~\cite{raichev}. This provides a useful analytical check of the finite-$b$ extension used here.

Current-penetrable contacts can additionally produce a magnetic-field-dependent jump of the electrochemical potential \cite{raichev}. The experimentally measured coefficient may therefore be written more generally as
\begin{equation}
\alpha_{\mathrm{exp}}(T)
=
\alpha_{\mathrm{bulk}}(T)
+
\alpha_{\mathrm{boun}}(T).
\label{eq:alphatotal}
\end{equation}
The first term is described by Eqs.~(\ref{eq:bulkMR})--(\ref{eq:phiSlip}), while the second originates from the boundary potential discontinuity. In the Stokes regime, Ref.~\cite{raichev} shows that this contribution is parametrically smaller than the bulk term by a factor of order $(l_{ee}/d)^2$, where $d=R_b-R_a$. Since a complete expression for this boundary term throughout the Stokes--Ohm crossover is not presently available, we do not introduce it as an additional fitting parameter. Instead, the finite-slip bulk solution is used for the quantitative extraction of $l_G$.
In the strict Stokes regime this boundary contribution is
parametrically smaller than the bulk term. Outside this limit, its
complete crossover form is not presently available, and we therefore
retain it as a separate systematic correction.

\section{Discussion and comparison with experiment}

The measured magnetoresistance is purely quadratic in the low-field range used in this work; therefore we write
\[
R(B,T)-R(0,T)=\alpha(T)B^2.
\]

The experimental coefficient $\alpha(T)$ is analyzed using Eq.~(\ref{eq:bulkMR}) with the finite-slip crossover function of Eq.~(\ref{eq:phiSlip}). In contrast to our previous parametrization, no effective terminal resistivity is introduced. The intrinsic bulk resistivity is calculated consistently from
\[
\rho_0(T)=
\frac{m^\ast}
{n_s e^2\tau_{\mathrm{mr}}(T)},
\]
while the same $\tau_{\mathrm{mr}}(T)$ enters
\[
l_G(T)=\sqrt{\nu(T)\tau_{\mathrm{mr}}(T)}.
\]
Thus, the momentum-relaxation contribution is treated consistently in both the Drude prefactor and the Stokes--Ohm crossover.

The measured two-terminal zero-field resistance is not used to determine $\rho_0(T)$, because it contains a much larger contact contribution. Instead, the bulk momentum-relaxation rate is parameterized as
\begin{equation}
\tau_{\mathrm{mr}}^{-1}(T)
=
\tau_0^{-1}
+
B_{\mathrm{ph}}T,
\label{eq:taumr}
\end{equation}
with common parameters constrained by the independently characterized bulk transport. The finite slip length is fixed at the value consistent with our Hall-bar devices rather than treated as an unconstrained fitting parameter.

After \(l_G(T)\) is obtained from \(\alpha(T)\), the relaxation time \(\tau_2\) can be estimated from
\[
\tau_2=
\frac{4l_G^2}
{v_F^2\tau_{\mathrm{mr}}}.
\]
For a GaAs quantum well with \(m^\ast=0.067m_e\) and
\(n_s=7.1\times10^{11}~\mathrm{cm^{-2}}\), we use $k_F=\sqrt{2\pi n_s}$ and $v_F=\frac{\hbar k_F}{m^\ast}$, giving \(v_F\simeq3.6\times10^5~\mathrm{m/s}\).

The relaxation of the second angular harmonic contains two contributions:
\begin{equation}
\tau_2^{-1}(T)
=
\tau_{2,\mathrm{imp}}^{-1}
+
\tau_{2,\mathrm{ee}}^{-1}(T),
\label{eq:tau2}
\end{equation}
where \(\tau_{2,\mathrm{imp}}^{-1}\) is the temperature-independent residual contribution and
\begin{equation}
\tau_{2,\mathrm{ee}}^{-1}(T)
=
A_{ee}
\frac{k_B^2}{\hbar E_F}
T^2
\label{eq:tau2ee}
\end{equation}
is the electron-electron contribution. Here \(E_F\) is the Fermi energy and \(A_{ee}\) is a dimensionless coefficient characterizing electron-electron scattering.

The analysis shown in Fig.~4 was performed using a constrained fitting
procedure. The $T^2$ electron--electron scattering contribution was
constrained to the range independently inferred from the Hall-bar
measurements, while the principal sample-dependent fitting parameter was
the residual relaxation rate $1/\tau_{2,\mathrm{imp}}$. The similar
$T^2$ slopes obtained for the different Corbino geometries are therefore
a consequence of this common electron--electron scattering scale.

Figure~4(a) shows the Gurzhi length extracted using the finite-slip
crossover. Despite the substantial difference in device dimensions, the
values obtained for different outer radii exhibit the same overall
temperature dependence. This is consistent with $l_G$ being primarily a
bulk transport length rather than a contact-specific quantity.

Finite slip is important for obtaining this consistent picture. For a
given measured magnetoresistance, finite slip increases the crossover
function relative to the no-slip solution. Consequently, the same
experimental value of $\alpha$ corresponds to a larger Gurzhi length and
a smaller residual second-harmonic relaxation rate. The resulting
$\tau_{2,\mathrm{imp}}^{-1}$ values are comparable among devices with
different radii and remain of the same order as those inferred from
mesoscopic Hall-bar measurements.

\begin{figure} 
 \includegraphics[width=9cm]{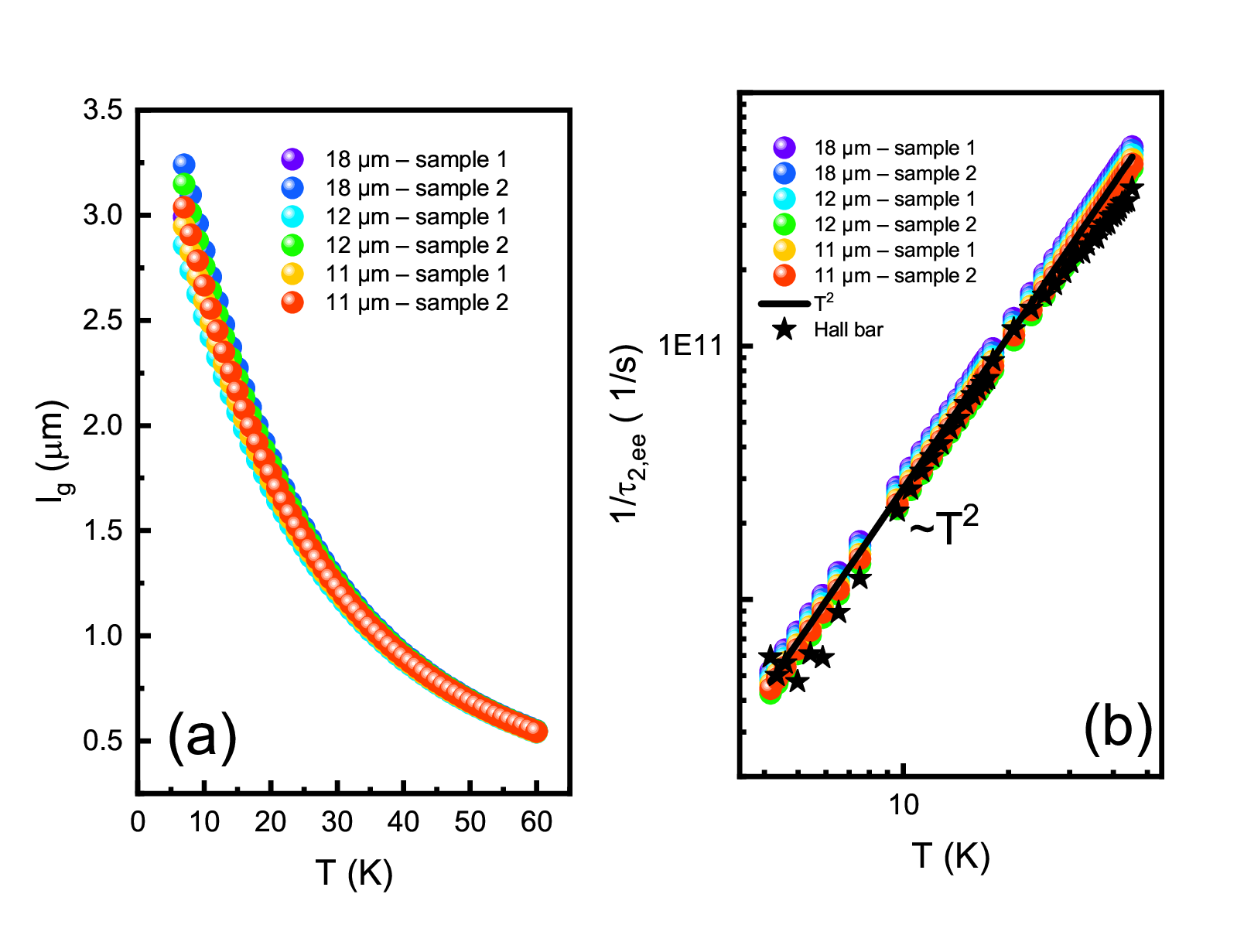}
\caption{(Color online) (a) Temperature dependence of the Gurzhi length \(l_G\) for Corbino devices with different outer radii. Two samples were studied for each device radius. The values are extracted using the finite-slip Stokes--Ohm crossover.} (b) Temperature dependence of the scattering rate \(1/\tau_{2,ee}\) for Corbino devices with different outer radii. Two samples were studied for each device radius. Stars denote \(1/\tau_{2,ee}\) values extracted from comparison with the theory of magnetoresistance measured in a narrow Hall bar; see Ref.~\cite{levin3}. The solid line shows a \(T^2\) dependence.
\end{figure}

The electron-electron contribution obtained after subtraction of the residual term is shown in Fig.~4(b). The six Corbino samples exhibit approximately quadratic temperature dependences, with moderate sample-to-sample variations of the coefficient $A_{ee}$.  The Hall-bar measurements provide an independent external constraint on the electron-electron scattering scale used in the finite-slip
analysis. The value of the slip length, $l_s \approx 4~\mu\mathrm{m}$, is comparable to the scale observed in Hall-bar samples (see, for example, Ref.~\cite{levin2}). At the same time, the slip length may vary between samples with different configurations and geometries \cite{gusev1, gusev2}, as expected from its sensitivity to the microscopic boundary conditions and details of sample fabrication. The apparent difference between Hall-bar and Corbino transport should
not be interpreted as evidence that the electron system is hydrodynamic
in one geometry but not in the other. The microscopic scattering
hierarchy is primarily a property of the electron system, whereas its
transport manifestation is strongly geometry dependent. In a Hall bar,
the lateral boundaries generate shear and can produce a pronounced
viscous correction to the longitudinal resistance. In a Corbino disk,
by contrast, the nearly potential radial flow couples much more weakly
to viscosity at zero magnetic field. The viscous response becomes
observable mainly through the magnetic-field-induced azimuthal flow
and the associated Stokes--Ohm crossover. Thus, even for the same
underlying electron-electron scattering scale, the magnitude and form
of the hydrodynamic correction can be substantially different in the
two geometries.

\begin{table}[ht]
\caption{\label{tab1}
Fitting parameters of the electron system for different Corbino outer radii.
The finite-slip boundary condition is included in the analysis.}
Parameters are defined in the text.
\begin{ruledtabular}
\begin{tabular}{lcccccc}
& $R_b$
& $1/\tau_{2,\mathrm{imp}}$
& $1/\tau_0$
& $A_{ee}$
& $B_{\mathrm{ph}}$
& $l_s$
\\
& $(\mu\mathrm{m})$
& $(10^{11}\,\mathrm{s}^{-1})$
& $(10^{10}\,\mathrm{s}^{-1})$
&
& $(10^{9}\,\mathrm{s}^{-1}\mathrm{K}^{-1})$
& $(\mu\mathrm{m})$
\\
\hline
& 11 & $2.32$ & $0.90$ & $0.6$ & $1.06$ & $4.0$ \\
& 12 & $2.22$ & $0.90$ & $0.6$ & $1.06$ & $4.0$ \\
& 18 & $2.10$ & $0.90$ & $0.66$ & $1.06$ & $4.0$ \\
\end{tabular}
\end{ruledtabular}
\end{table}

For completeness, the finite-slip theory also allows a direct extrapolation of the measured system toward the two ideal limits requested for comparison. No additional fitting parameters are required for this extrapolation. Taking $b\rightarrow0$ while retaining the same viscosity gives the pure hydrodynamic response of Eq.~(\ref{eq:hydroLimit}), whereas taking $b\rightarrow\infty$ gives the purely diffusive Drude result of Eq.~(\ref{eq:diffLimit}). The experimental devices occupy the intermediate Stokes--Ohm crossover rather than either asymptotic limit.

This comparison is important for the interpretation of positive Corbino magnetoresistance. A quadratic positive magnetoresistance alone is not a unique signature of hydrodynamics, because the purely diffusive Corbino geometry also produces a $B^2$ response. The hydrodynamic information is contained instead in the deviation from the diffusive limit and in its systematic evolution with the dimensionless ratio $R_b/l_G$. At low temperature the larger Gurzhi length enhances the viscous correction, whereas increasing temperature reduces $l_G$ and continuously drives the response toward the diffusive limit.

The finite-slip solution therefore provides a single description of the complete evolution
\[
\text{Stokes}
\quad\longrightarrow\quad
\text{Stokes--Ohm crossover}
\quad\longrightarrow\quad
\text{diffusive},
\]
while reproducing the known analytical expressions in both asymptotic limits.

Finally, current-penetrable contacts can generate the additional field-dependent boundary voltage discussed in Ref.~\cite{raichev}. This effect does not invalidate the bulk analysis, but it sets the natural accuracy with which the viscous parameters can be extracted from a two-terminal Corbino device. The good agreement between different ring geometries and with the independent Hall-bar electron-electron scattering rate indicates that the dominant temperature-dependent magnetoresistance is captured by the bulk finite-slip crossover. The remaining boundary contribution should therefore be regarded as a correction rather than the origin of the observed effect.

\section{Conclusions}

We have measured the magnetoresistance of high-mobility GaAs Corbino
devices and observed a pronounced positive quadratic dependence on
perpendicular magnetic field over a broad intermediate-temperature
range. Since the classical Drude response of a Corbino disk is itself
quadratic in magnetic field, the observation of positive
magnetoresistance alone does not uniquely identify hydrodynamic
transport. The hydrodynamic interpretation is instead
based on the systematic temperature and size dependence of the
magnetoresistance within the Stokes--Ohm crossover.

By comparing the data with the Corbino Stokes--Ohm theory,
extended here to finite-slip boundary conditions at
current-penetrable contacts, we extracted the temperature dependence
of the Gurzhi length and the relaxation rate of the second angular
harmonic. The Hall-bar measurements provide an
external constraint on the electron-electron scattering scale used in
this analysis. Within this constrained range, the extracted Gurzhi
length decreases monotonically with temperature, the residual
second-harmonic relaxation rates are comparable for different Corbino
radii, and the electron-electron contribution remains approximately
proportional to $T^2$.

The finite-slip Stokes--Ohm solution continuously
connects the two limiting regimes. In the hydrodynamic Stokes limit,
the momentum-relaxing Drude resistivity drops out and the
magnetoresistance is controlled directly by the viscosity, whereas in
the diffusive limit the conventional Drude Corbino response is
recovered. The experimental devices lie in the intermediate crossover
between these limits. The comparison between different ring radii is
consistently described by the bulk finite-slip crossover. At the same
time, current-penetrable contacts remain relevant: finite slip modifies
the crossover itself, while the magnetic-field-dependent
electrochemical-potential jump provides a separate boundary correction
\cite{raichev}.

The Corbino geometry therefore provides a
complementary probe to Hall-bar measurements. The absence of extended
side boundaries reduces the ambiguity associated with edge-flow
boundary conditions, while the Hall-bar measurements constrain the
bulk momentum-relaxation and electron-electron scattering scales. Our
results show that the combination of these two geometries provides a
consistent description of viscous electron transport in high-mobility
GaAs two-dimensional electron systems. Further experiments on smaller
and narrower Corbino structures should enhance the sensitivity to
boundary effects, while multiterminal Corbino devices could help
separate the series-contact contribution, provided that the additional
concentric interfaces are included in the corresponding multiboundary
hydrodynamic treatment.

\section {Acknowledgments}
This work is supported by FAPESP (São Paulo Research Foundation) Grants No. 2019/16736-2, No. 2021/12470- 8, No. 2025/22786-3, CNPq (National Council for Scientific and Technological Development). The growth of GaAs quantum wells and preliminary transport measurements were supported by the Russian Science Foundation (Grant No. 23-72-30003). L.A. Semenova is acknowledged for performing the photolithographic processing.
\section{DATA AVAILABILITY}
The data that support the findings of this article are openly
available \cite{zenodo}.


\clearpage
\pagebreak
\onecolumngrid

\setcounter{equation}{0}
\setcounter{figure}{0}
\setcounter{table}{0}
\setcounter{page}{1}
\setcounter{section}{0}

\makeatletter
\renewcommand{\theequation}{S\arabic{equation}}
\renewcommand{\thefigure}{S\arabic{figure}}
\renewcommand{\thetable}{S\arabic{table}}
\renewcommand{\thepage}{S\arabic{page}}
\makeatother

\begin{center}
  \textbf{\large Supplemental Material for ``Hydrodynamic magnetotransport in a GaAs Corbino geometry''}\\[1em]
  
  \normalsize A. D. Levin,$^1$ A. S. Jaroshevich,$^{2}$ Z. D. Kvon,$^{2,3}$
  V. A. Chitta,$^1$ M. S. Aksenov,$^{2}$ D. V. Dmitriev,$^{2}$
  A. K. Bakarov$^{2}$ and G. M. Gusev$^1$\\[0.5em]
  
  \small
  \textit{$^1$Instituto de F\'{\i}sica da Universidade de S\~ao Paulo, S\~ao Paulo, SP, Brazil}\\
  \textit{$^2$Institute of Semiconductor Physics, Novosibirsk 630090, Russia}\\
  \textit{$^3$Novosibirsk State University, Novosibirsk 630090, Russia}
  
  \vspace{0.5em}
  (\today)
\end{center}

\vspace{1em}

\begin{quotation}
\noindent\small
This Supplemental Material presents the derivation and numerical checks
of the finite-slip Stokes--Ohm crossover used in the main text. We verify
its no-slip, hydrodynamic, and diffusive limits, compare the measured
magnetoresistance with the Drude baseline, and examine the stability of
the constrained fit. We also distinguish the finite-slip modification
of the bulk response from the separate field-dependent boundary
contribution.
\end{quotation}

\vspace{1.5em}

\section{Finite-slip Stokes--Ohm problem}

Charge conservation fixes the radial current density in an annulus,
\begin{equation}
j_r(r)=\frac{I}{2\pi r}.
\label{eq:jr}
\end{equation}
In a perpendicular magnetic field, the Hall force generates an
azimuthal current $j_\phi(r)$. In the linear Stokes--Ohm description it
obeys
\begin{equation}
l_G^2\left[
\frac{d^2 j_\phi}{dr^2}
+\frac{1}{r}\frac{d j_\phi}{dr}
-\frac{j_\phi}{r^2}
\right]
-j_\phi
=
\omega_c\tau_{\rm mr}\,j_r(r),
\label{eq:stokesohm}
\end{equation}
where
\begin{equation}
l_G=\sqrt{\nu\tau_{\rm mr}},
\qquad
\nu=\frac{v_F^2\tau_2}{4},
\qquad
\omega_c=\frac{eB}{m^\ast}.
\end{equation}
The bulk sheet resistivity is
\begin{equation}
\rho_0(T)=\frac{m^\ast}{n_s e^2\tau_{\rm mr}(T)}.
\label{eq:rho0}
\end{equation}

Introducing $x=r/l_G$, the general solution of
Eq.~(\ref{eq:stokesohm}) can be written as
\begin{equation}
j_\phi(r)=
\frac{IB}{2\pi\rho_0 n_s e\,l_G}
\left[
C_I I_1(x)+C_KK_1(x)-\frac{1}{x}
\right],
\label{eq:jphi}
\end{equation}
where $I_n$ and $K_n$ are modified Bessel functions.

For current-penetrable contacts we impose Navier slip boundary
conditions,
\begin{equation}
j_\phi(R_a)
=
l_s
\left.\frac{dj_\phi}{dr}\right|_{R_a},
\qquad
j_\phi(R_b)
=
-l_s
\left.\frac{dj_\phi}{dr}\right|_{R_b}.
\label{eq:slipbc}
\end{equation}
The opposite signs follow from the opposite directions of the inward
normal at the inner and outer boundaries.

We define
\begin{equation}
a=\frac{R_b}{R_a},
\qquad
b=\frac{R_b}{l_G},
\qquad
x_a=\frac{b}{a},
\qquad
x_b=b,
\qquad
\lambda=\frac{l_s}{l_G}.
\end{equation}
Using
\[
I_1'(x)=I_0(x)-\frac{I_1(x)}{x},
\qquad
K_1'(x)=-K_0(x)-\frac{K_1(x)}{x},
\]
the two coefficients in Eq.~(\ref{eq:jphi}) are obtained from
\begin{widetext}
\begin{equation}
\begin{pmatrix}
I_1(x_a)-\lambda I_1'(x_a) &
K_1(x_a)-\lambda K_1'(x_a)
\\
I_1(x_b)+\lambda I_1'(x_b) &
K_1(x_b)+\lambda K_1'(x_b)
\end{pmatrix}
\begin{pmatrix}
C_I\\
C_K
\end{pmatrix}
=
\begin{pmatrix}
x_a^{-1}+\lambda x_a^{-2}\\
x_b^{-1}-\lambda x_b^{-2}
\end{pmatrix}.
\label{eq:matrix}
\end{equation}
\end{widetext}
Integrating the radial electric field across the annulus gives
\begin{equation}
\Delta R_{\rm bulk}(B,T)
=
\frac{B^2\ln a}
{2\pi\rho_0(T)(n_s e)^2}
\,
\Phi_{\rm slip}(a,b,\lambda),
\label{eq:bulk}
\end{equation}
where we define the physical crossover multiplier directly as
\begin{equation}
\Phi_{\rm slip}
=
1-
\frac{
C_I\left[I_0(b)-I_0(b/a)\right]
+
C_K\left[K_0(b/a)-K_0(b)\right]
}
{\ln a}.
\label{eq:phi}
\end{equation}
This notation avoids the $f$ versus $1-f$ convention ambiguity:
$\Phi_{\rm slip}$ is the quantity that multiplies the Drude Corbino
magnetoresistance. It satisfies
\begin{equation}
\Phi_{\rm slip}\rightarrow0
\quad (b\rightarrow0),
\qquad
\Phi_{\rm slip}\rightarrow1
\quad (b\rightarrow\infty).
\end{equation}

\begin{figure}[t]
\centering
\includegraphics[width=0.96\columnwidth]{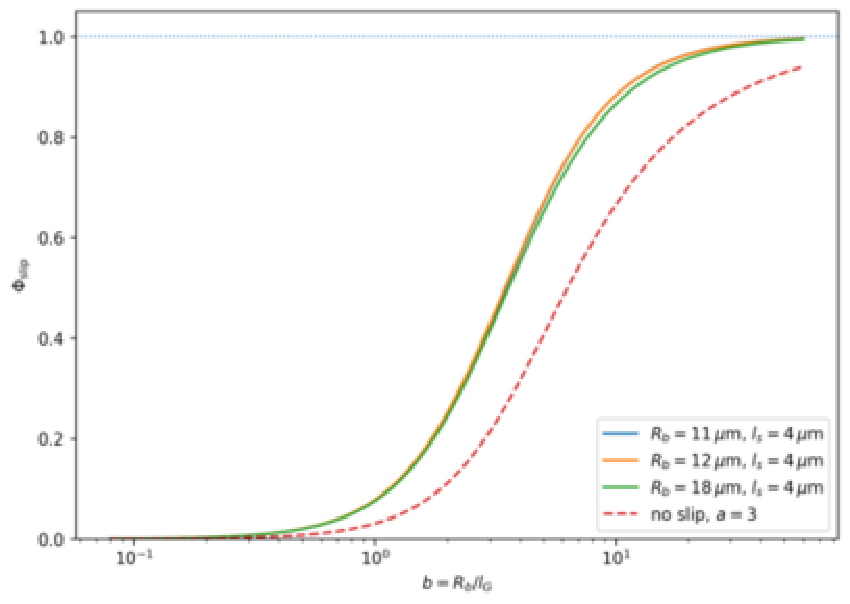}
\caption{
Finite-slip Stokes--Ohm crossover multiplier
$\Phi_{\rm slip}$ as a function of $b=R_b/l_G$ for the three Corbino
geometries, using $R_a=6~\mu$m and $l_s=4~\mu$m.  A representative
no-slip curve is shown for comparison.  All curves approach the
diffusive value $\Phi=1$ at large $b$.
}
\label{fig:S1}
\end{figure}

For $l_s=0$, Eq.~(\ref{eq:matrix}) becomes the no-slip boundary problem
and Eq.~(\ref{eq:phi}) reduces to the Stokes--Ohm crossover of
Ref.~\cite{levchenko2}. Figure~\ref{fig:S1} shows the finite-slip
crossover for the three Corbino geometries used experimentally. A
finite slip length increases $\Phi$ at fixed $b$ and therefore shifts
the Gurzhi length extracted from a given measured magnetoresistance to
larger values.

\section{Pure hydrodynamic and pure diffusive limits}

The same finite-slip solution continuously connects the two limiting
transport regimes. This is useful because a positive $B^2$
magnetoresistance is present even in the purely diffusive Corbino
geometry; the viscous information is contained in the deviation from
that limit.

\subsection{Diffusive limit}

For
\begin{equation}
b=\frac{R_b}{l_G}\gg1,
\end{equation}
the viscous boundary layer is narrow compared with the device
dimensions and
\begin{equation}
\Phi_{\rm slip}\rightarrow1.
\end{equation}
Equation~(\ref{eq:bulk}) becomes
\begin{equation}
\Delta R_{\rm diff}
=
\frac{B^2\ln a}
{2\pi\rho_0(T)(n_s e)^2}.
\label{eq:diff}
\end{equation}
Using
\[
R_0=\frac{\rho_0}{2\pi}\ln a,
\qquad
\mu=\frac{e\tau_{\rm mr}}{m^\ast},
\]
this is the conventional Drude Corbino result
\begin{equation}
\Delta R_{\rm diff}=R_0(\mu B)^2.
\label{eq:drude}
\end{equation}

\subsection{Hydrodynamic Stokes limit}

For
\begin{equation}
b\ll1,
\end{equation}
the exact finite-slip kernel has the expansion
\begin{equation}
\Phi_{\rm slip}
\simeq
\frac{b^2(1-a^{-2})}
{8\ln a}
G_{\rm corb}(l_s).
\label{eq:stokesphi}
\end{equation}
Substitution into Eq.~(\ref{eq:bulk}), together with
$l_G^2=\nu\tau_{\rm mr}$ and
$\rho_0=m^\ast/(n_s e^2\tau_{\rm mr})$, cancels the
momentum-relaxation time exactly and gives, in SI units,
\begin{equation}
\Delta R_{\rm Stokes}
=
\frac{B^2(R_b^2-R_a^2)}
{16\pi\eta}
G_{\rm corb},
\qquad
\eta=m^\ast n_s\nu.
\label{eq:stokesR}
\end{equation}
Thus the purely hydrodynamic magnetoresistance is controlled directly
by the viscosity and is independent of the Drude momentum-relaxation
time.

For completeness, the geometrical factor $G_{\rm corb}$ obtained by
Raichev for finite slip \cite{raichev} is
\begin{equation}
G_{\rm corb}
=
1-2c_0
-
2(1+2c_1)
\frac{R_b^2}{R_b^2-R_a^2}
\ln\frac{R_b}{R_a},
\label{eq:Gcorb}
\end{equation}
with
\begin{align}
c_0={}&-\frac{1}{D}\Bigg[
l_s\left(
\frac{R_b-l_s}{R_b^2}
+
\frac{R_a+l_s}{R_a^2}
\right)
\nonumber\\
&\hspace{1.7cm}
+
\frac{(R_b+l_s)(R_a+l_s)}{R_a^2}
\ln\frac{R_b}{R_a}
\Bigg],
\label{eq:c0}\\
c_1={}&\frac{1}{D}\Bigg[
l_s\frac{R_b+R_a}{R_b^2}
+
\frac{(R_b+l_s)(R_a-l_s)}{R_b^2}
\ln\frac{R_b}{R_a}
\Bigg],
\label{eq:c1}\\
D={}&
\frac{(R_b+l_s)(R_a+l_s)}{R_a^2}
-
\frac{(R_b-l_s)(R_a-l_s)}{R_b^2}.
\label{eq:D}
\end{align}

Equations~(\ref{eq:Gcorb})--(\ref{eq:D}) are the Appendix-B result of
Ref.~\cite{raichev}. We used them as an independent check of the
finite-$b$ solution. For the actual three geometries and $l_s=4~\mu$m,
evaluation at $b=10^{-3}$ gives agreement between
$\Phi_{\rm slip}/b^2$ and Eq.~(\ref{eq:stokesphi}) to better than
$1.1\times10^{-7}$ in relative accuracy.

\begin{figure}[t]
\centering
\includegraphics[width=0.96\columnwidth]{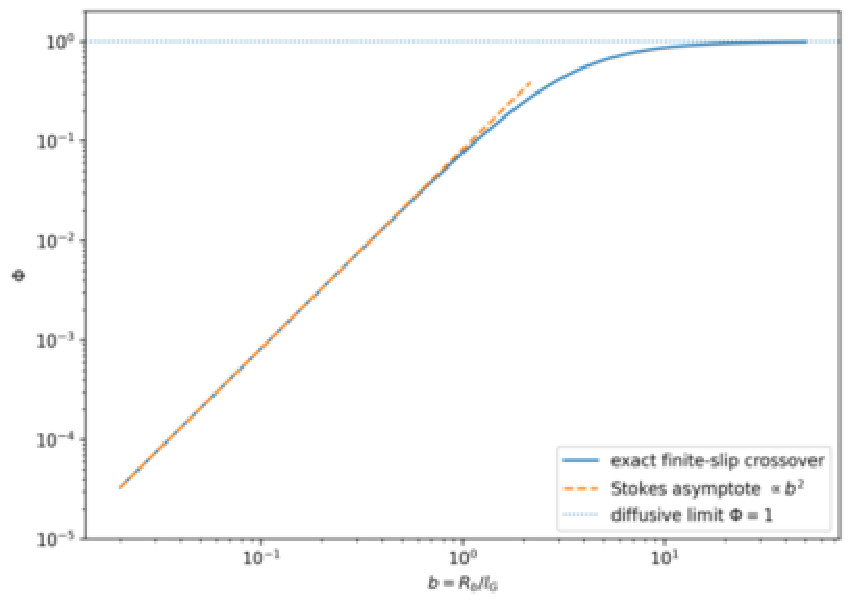}
\caption{
Exact finite-slip crossover for $R_a=6~\mu$m,
$R_b=18~\mu$m and $l_s=4~\mu$m.  The small-$b$ line is the
finite-slip Stokes asymptote of Ref.~\cite{raichev}; the horizontal
line is the purely diffusive limit.  The same solution continuously
connects the two regimes.
}
\label{fig:S2}
\end{figure}

Figure~\ref{fig:S2} shows the full finite-slip crossover together with
the Stokes expansion and the pure diffusive limit for a representative
$R_b=18~\mu$m ring. The Stokes expression is recovered at small $b$,
whereas the exact solution saturates to the Drude value at large $b$.

\subsection{Comparison with the bulk Drude baseline}

It is useful to compare the measured quadratic magnetoresistance with
the conventional Corbino Drude response calculated from the independently
characterized bulk transport. The corresponding coefficient is
\begin{equation}
\alpha_{\rm Drude}^{\rm bulk}(T)
=
\frac{\ln(R_b/R_a)}
{2\pi\rho_0(T)(n_s e)^2},
\label{eq:alphaDrude}
\end{equation}
where $\rho_0(T)$ is obtained from the bulk momentum-relaxation rate
inferred from the macroscopic Hall-bar measurements.

This comparison, however, is not by itself a unique test of
hydrodynamic transport. Within the Drude model,
\begin{equation}
\alpha_{\rm Drude}
=
\frac{\mu}{2\pi n_s e}
\ln\frac{R_b}{R_a},
\end{equation}
and therefore any positive quadratic magnetoresistance coefficient can
formally be reproduced by introducing the effective mobility
\begin{equation}
\mu_{\rm req}^{\rm Drude}(T)
=
\frac{2\pi n_s e}
{\ln(R_b/R_a)}
\alpha_{\rm exp}(T).
\label{eq:mureq}
\end{equation}
The experimentally measured ratio can thus be written as
\begin{equation}
\frac{\alpha_{\rm exp}}
{\alpha_{\rm Drude}^{\rm bulk}}
=
\frac{\mu_{\rm req}^{\rm Drude}}
{\mu_{\rm bulk}}.
\label{eq:ratioDrude}
\end{equation}

\begin{figure}[t]
\centering
\includegraphics[width=0.96\columnwidth]
{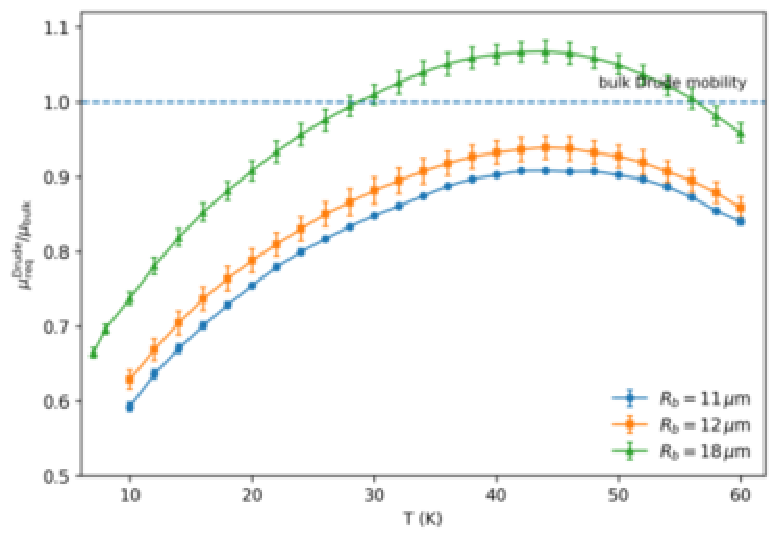}
\caption{
Effective mobility required to reproduce the measured quadratic
Corbino magnetoresistance within a purely Drude description, normalized
by the independently characterized bulk mobility. By construction,
$\mu_{\rm req}^{\rm Drude}/\mu_{\rm bulk}
=\alpha_{\rm exp}/\alpha_{\rm Drude}^{\rm bulk}$.
Symbols show the mean of the two nominally identical devices for each
outer radius, and the error bars indicate half of their
sample-to-sample spread. The dashed line corresponds to the bulk Drude
mobility. Deviations from unity quantify the change in effective
mobility that would be required by a purely Drude interpretation; they
do not, by themselves, constitute evidence for hydrodynamic transport.
}
\label{fig:S3drude}
\end{figure}

Figure~\ref{fig:S3drude} shows this quantity for the three Corbino
geometries. At low temperature, reproducing the measured
magnetoresistance within a purely Drude description would require an
effective mobility smaller than the macroscopic bulk value. Such a
reduction is not unexpected in a mesoscopic device, since finite-size
scattering, boundaries, and contact regions can modify the effective
momentum-relaxation rate. Conversely, a modest ratio above unity, as
observed for the largest ring over part of the temperature range, may
reflect sample-to-sample variation of the bulk transport parameters or
additional contributions to the measured magnetoresistance.

Consequently, Fig.~\ref{fig:S3drude} should not be interpreted as an
independent demonstration of hydrodynamic transport. Its purpose is to
show explicitly how far the measured response lies from the Drude
baseline constructed using the independently characterized bulk
mobility, and equivalently how much the effective mobility would have
to be modified in a purely Drude interpretation. The hydrodynamic
analysis imposes a more restrictive condition: the independently
constrained bulk momentum-relaxation scale is retained, while the
temperature and size dependence of the magnetoresistance is described
by the finite-slip Stokes--Ohm crossover. The viscous relaxation scale
obtained in this way can then be compared with that inferred
independently from Hall-bar measurements.

\section{Role of current-penetrable contacts}

The large zero-field two-terminal contact resistance does not replace
the bulk sheet resistivity in Eq.~(\ref{eq:bulk}).  The measured
resistance can be separated as
\begin{equation}
R_{\rm meas}(B,T)
=
R_c(0,T)+R_{\rm bulk}(0,T)
+
\left[
\alpha_{\rm bulk}(T)+\alpha_{\rm boun}(T)
\right]B^2
+O(B^4).
\label{eq:Rmeas}
\end{equation}
Therefore the field-independent contact resistance $R_c(0,T)$ cancels
from $R(B,T)-R(0,T)$.  The resistivity entering
Eq.~(\ref{eq:bulk}) remains the intrinsic bulk
$\rho_0(T)$ of Eq.~(\ref{eq:rho0}).

There are nevertheless two genuine contact effects.  First, the
tangential current satisfies the finite-slip condition
Eq.~(\ref{eq:slipbc}), which modifies the \emph{bulk} crossover
function.  Second, a current-penetrable boundary has an
electrochemical-potential jump.  Raichev showed that this jump acquires
a magnetic-field-dependent contribution proportional to the normal
derivative of the tangential current \cite{raichev}; this gives the
additive coefficient $\alpha_{\rm boun}$ in Eq.~(\ref{eq:Rmeas}).

In the strict Stokes regime, Ref.~\cite{raichev} finds that the
field-dependent boundary resistance is smaller than the bulk
magnetoresistance by a factor of order
\begin{equation}
\left(\frac{l_{ee}}{d}\right)^2,
\qquad
d=R_b-R_a.
\end{equation}
Our devices are in the Stokes--Ohm crossover rather than in the strict
$d\ll l_G$ limit, so we do not use this estimate as a quantitative
fitting formula.  The boundary magnetoresistance is therefore treated
as a systematic correction, while the finite-slip bulk crossover is
included explicitly.

The slip length itself can depend on magnetic field
\cite{raichev}.  This does not affect the extraction of the
\emph{quadratic} low-field coefficient to leading order.  Time-reversal
symmetry gives
\begin{equation}
l_s(B)=l_s(0)+O(B^2),
\end{equation}
and hence
\[
B^2\Phi[l_s(B)]
=
B^2\Phi[l_s(0)]
+
O(B^4).
\]
Consequently, the zero-field slip length is the appropriate quantity
in the coefficient $\alpha$ used here.

\section{Constrained fit and stability}

The momentum-relaxation rate is parameterized as
\begin{equation}
\tau_{\rm mr}^{-1}(T)
=
\tau_0^{-1}
+
B_{\rm ph}T,
\label{eq:taumr}
\end{equation}
and the second-harmonic relaxation rate as
\begin{equation}
\tau_2^{-1}(T)
=
\tau_{2,\rm imp}^{-1}
+
A_{ee}
\frac{k_B^2T^2}{\hbar E_F}.
\label{eq:tau2}
\end{equation}
The same $\tau_{\rm mr}(T)$ is used in both the Drude resistivity,
Eq.~(\ref{eq:rho0}), and
$l_G=(v_F/2)\sqrt{\tau_2\tau_{\rm mr}}$.  This is required by the
Stokes--Ohm derivation; in the Stokes limit the explicit
$\tau_{\rm mr}$ dependences then cancel automatically.

For the robustness test, the low-temperature bulk resistivity is
anchored at
\begin{equation}
\rho_0(4.2~{\rm K})=4.5~\Omega,
\end{equation}
and $\tau_0^{-1}$ is adjusted consistently for each trial
$B_{\rm ph}$.  We scan fixed slip lengths
\[
l_s=2,\;2.5,\;3,\;3.5,\;4~\mu{\rm m}
\]
and fit the six measured $\alpha(T)$ data sets simultaneously.  We
then identify the region in which the following physically motivated
conditions hold:
\begin{equation}
0.6\le A_{ee}\le1,
\end{equation}
\begin{equation}
1\times10^{11}
\le
\tau_{2,\rm imp}^{-1}
\le
3\times10^{11}~{\rm s}^{-1},
\end{equation}
and
\begin{equation}
2~\mu{\rm m}\le l_G(8~{\rm K})\le3~\mu{\rm m}.
\end{equation}

\begin{figure}[t]
\centering
\includegraphics[width=0.96\columnwidth]{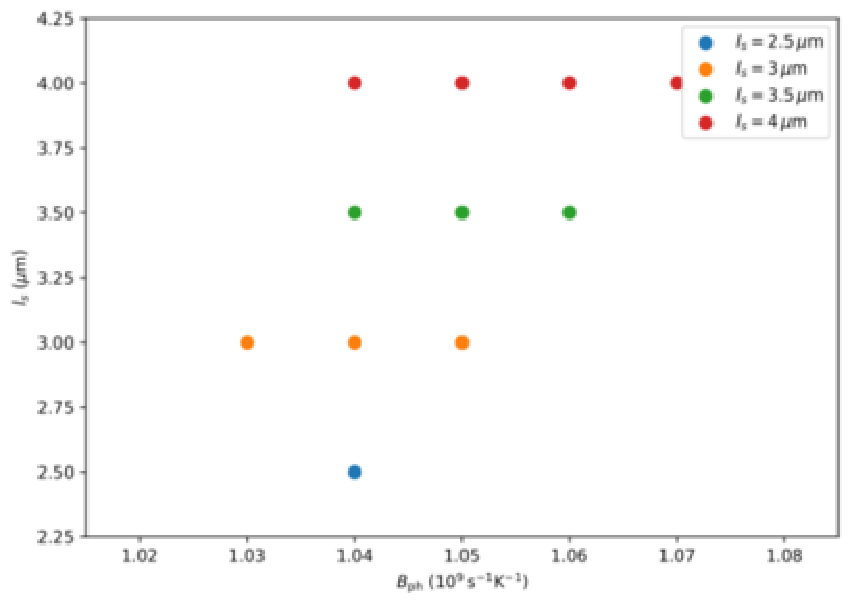}
\caption{
Region of the constrained fit satisfying simultaneously
$0.6\le A_{ee}\le1$,
$1\le\tau_{2,\rm imp}^{-1}/10^{11}{\rm s}^{-1}\le3$, and
$2\le l_G(8\,{\rm K})\le3~\mu$m.
The physically acceptable solutions form an extended region for
$l_s\simeq3$--$4~\mu$m.
}
\label{fig:S3}
\end{figure}

The accepted points are shown in Fig.~\ref{fig:S3} and summarized in
Table~\ref{tab:window}.  The important result is that the physical
solution occupies a finite region of parameter space for
$l_s\simeq3$--$4~\mu$m rather than a single isolated combination of
parameters.  This range of slip lengths is also of the same order as
values independently inferred in our Hall-bar devices, although the
microscopic boundary conditions of a Hall-bar edge and a
current-penetrable Corbino contact need not be identical.

\begin{table}[t]
\caption{
Ranges of the constrained solutions satisfying all three physical
criteria.  $B_{\rm ph}$ is in
$10^9~{\rm s}^{-1}{\rm K}^{-1}$,
$\tau_{2,\rm imp}^{-1}$ in $10^{11}~{\rm s}^{-1}$, and lengths are in
$\mu$m.
}
\label{tab:window}
\begin{ruledtabular}
\begin{tabular}{ccccc}
$l_s$ & $B_{\rm ph}$ & $A_{ee}$ &
$1/\tau_{2,\rm imp}$ & $l_G(8\,{\rm K})$\\
\hline
2.5 & 1.04 & 0.88 & 2.92 & 2.46\\
3.0 & 1.03--1.05 & 0.65--0.90 & 2.52--2.68 & 2.59--2.62\\
3.5 & 1.04--1.06 & 0.69--0.93 & 2.21--2.36 & 2.74--2.77\\
4.0 & 1.04--1.07 & 0.62--0.96 & 1.97--2.16 & 2.86--2.91\\
\end{tabular}
\end{ruledtabular}
\end{table}

\begin{figure}[t]
\centering
\includegraphics[width=0.96\columnwidth]{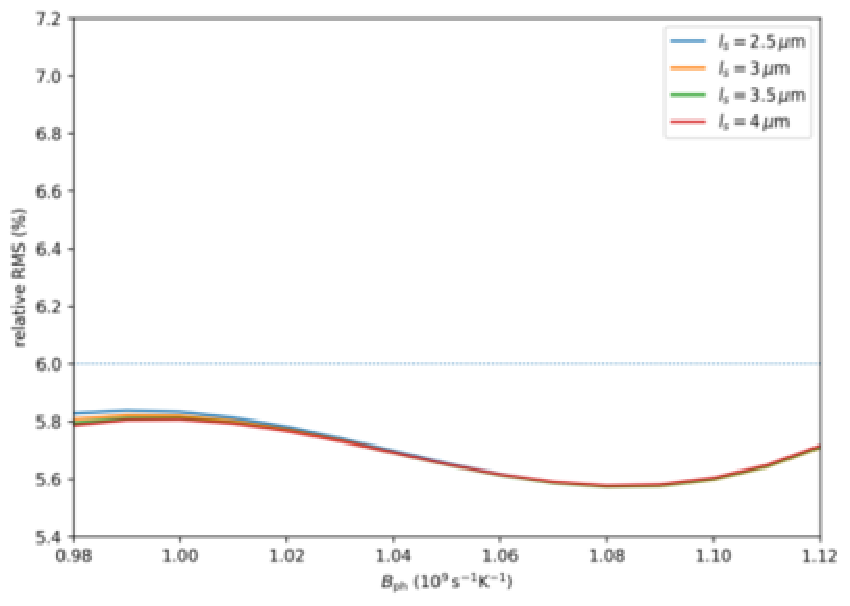}
\caption{
Relative RMS of the simultaneous fit to the six Corbino data sets as
a function of $B_{\rm ph}$ for several fixed slip lengths.  The fit
quality varies only weakly throughout the physically acceptable
region.
}
\label{fig:S4}
\end{figure}

Figure~\ref{fig:S4} shows the relative RMS of the global fit around
this region.  The minimum is broad: restricting the parameters to the
physical window changes the fit quality only weakly.  For example,
with $l_s=4~\mu$m, the solutions around
$B_{\rm ph}=1.04$--$1.07\times10^9~{\rm s}^{-1}{\rm K}^{-1}$
remain at a relative RMS of approximately $5.6\%$.

For the main-text analysis we use $l_s=4~\mu$m as a representative
finite-slip value.  At the central constrained point
$B_{\rm ph}=1.06\times10^9~{\rm s}^{-1}{\rm K}^{-1}$ the corresponding
residual momentum-relaxation rate fixed by the
$4.2$-K resistivity anchor is $\tau_0^{-1}\simeq0.90\times10^{10}~{\rm s}^{-1}$.
We emphasize that this simultaneous global scan is used only as
a robustness test and is distinct from the radius-resolved
constrained parameter extraction reported in the main text.
The extracted Gurzhi lengths are of order a few micrometers at low
temperature and decrease continuously toward the submicrometer scale
at high temperature, placing the measured devices in the intermediate
Stokes--Ohm crossover.

\end{document}